\documentclass[aps,preprint,eqsecnum,nofootinbib,amsmath,amssymb]{revtex4}

\newcommand\mystartdate{18 November 2007}

\newcommand\mydates{\mystartdate}

\newcommand{\smGT}{{\scriptscriptstyle >}}
\newcommand{\smLT}{{\scriptscriptstyle <}}

\newcommand{\smDelQ}{{\rm\scriptscriptstyle Q}}
\newcommand{\smBPS}{{\rm\scriptscriptstyle BPS}}
\newcommand{\smLIP}{{\rm\scriptscriptstyle LIP}}
\newcommand{\smCS}{{\rm\scriptscriptstyle CS}}
\newcommand{\smR}{{\rm\scriptscriptstyle R}}
\newcommand{\smS}{{\rm\scriptscriptstyle S}}
\newcommand{\smC}{{\rm\scriptscriptstyle C}}
\newcommand{\smQ}{{\rm\scriptscriptstyle Q}}
\newcommand{\smI}{{\rm\scriptscriptstyle I}}
\newcommand{\smM}{{\rm\scriptscriptstyle M}}
\newcommand{\smD}{{\rm\scriptscriptstyle D}}

\usepackage{graphicx} 
\usepackage{dcolumn}  
\usepackage{bm}       
\usepackage{latexsym} 
\usepackage{fancyhdr} 

\newcommand{\titleskip}{\baselineskip 18pt plus 1pt minus 1pt}
\newcommand{\affiliationskip}{\baselineskip 15pt plus 1pt minus 1pt}
\newcommand{\abstractskip}{\baselineskip 13pt plus 1pt minus 1pt}

\newcommand{\bodyskip}{\baselineskip 18pt plus 1pt minus 1pt}
\newcommand{\footnoteskip}{\baselineskip 12pt plus 1pt minus 1pt}

\newcommand{\bibliographyskip}{\baselineskip14pt plus 1pt minus 1pt}

\begin{document}

\preprint{\hfill LA-UR-07-7021~~Phys. Plasmas {\bf 15} (2008) 056302}

\title{\titleskip
Charged Particle Stopping Power Effects on Ignition: \\
Some Results from an Exact Calculation
}

\author{Robert L. Singleton Jr.}

\vskip0.2cm 
\affiliation{\affiliationskip
     Los Alamos National Laboratory\\
     Los Alamos, New Mexico 87545, USA
}
\date{\mydates}

\begin{abstract}
\abstractskip
\vskip0.3cm 

\noindent
A completely rigorous first-principles calculation of the charged
particle stopping power has recently been performed by Brown, Preston,
and Singleton (BPS). This calculation is exact to leading and
next-to-leading order in the plasma number density, including an exact
treatment of two-body quantum scattering. The BPS calculation is
therefore extremely accurate in the plasma regime realized during the
ignition and burn of an inertial confinement fusion capsule.
For deuterium-tritium fusion, the 3.5~MeV alpha particle range tends
to be 20--30\% longer than most models in the literature have
predicted, and the energy deposition into the ions tends to be
smaller. Preliminary numerical simulations indicate that this
increases the $\rho R$ required to achieve ignition.

\end{abstract}

\maketitle

\pagebreak
\bodyskip
\setcounter{page}{2}

\section{\label{intro} Introduction}

The charged particle stopping power of a hot plasma plays a critical
role in whether an inertial confinement fusion (ICF) capsule will
undergo thermonuclear burn and ignition. For a robust experimental
setup, in which the laser energy is sufficiently high to assure
ignition and full burn, the fine details of the stopping power are not
so relevant; however, in more marginal settings on the threshold of
ignition, these details are likely to play a more important role.
Conventional wisdom states that our current knowledge of the stopping
power is probably ``good enough''; however, I will present some
analytic and numerical results to illustrate that the subleading order
physics neglected or missed by most stopping power models, the terms
of order unity inside the Coulomb logarithm, can in fact lead to a
noticeable effect on ignition.

In this study, I would like to explore some implications of the
stopping power calculation of Brown, Preston, and Singleton
(BPS)~\cite{bps}. This calculation of the charged particle stopping
power, which includes the energy splitting of the projectile as it
slows down and a rigorous treatment of the quantum to classical
transition, is near-exact for the weakly coupled plasmas relevant for
ICF ignition. I should also emphasize that the BPS stopping power is
not a model of the Coulomb energy exchange, but a controlled first
principles calculation of this process. For a detailed pedagogical
explanation of the technique, please see Refs.~\cite{lfirst, bs,
bpsped}.  We may therefore study the effects of stopping power on
ignition and burn in a systematic fashion, and unlike utilizing a
model, we know the range of validity of the BPS stopping power, along
with an estimate of the error in any given plasma regime.  Rather than
launching straightway into the simulation of real ICF capsules, with
their concomitant computational and physical complexities, I shall
instead start more systematically and investigate the simpler problem
of the deuterium-tritium (DT) microsphere studied by Fraley {\em et
al.}~\cite{fraley74}.  We shall find that, for these so called Fraley
spheres, the BPS stopping power increases the ignition threshold
significantly compared to typical models of the stopping power found
in the literature. For example, the value of $\rho R$ necessary to
attain a self-sustaining burn in a Fraley sphere increases by 10\% in
comparison to the well-known state-of-the-art model of Li and
Petrasso~\cite{lip}.

Specializing to the case of DT fusion, the dominant boot-strap heating
mechanism necessary to achieve a self-sustained burn is $\alpha$
particle energy deposition in the background plasma. All things being
equal, the larger the stopping power $dE/dx$, the greater the energy
deposition of the supra-thermal 3.5\,MeV $\alpha$ particles, and
therefore the easier it becomes to achieve a self-sustained burn
through $\alpha$ particle heating of the ions. In other words, a
larger stopping power leads to a more efficient self-heating, and
conversely, a smaller charged particle stopping power would tend to
make ignition harder, since this would increase the range of $\alpha$
particles and lead to smaller energy deposition per unit volume.
Other factors, such as the relative amount of energy that the $\alpha$
particle apportions between plasma electrons and ions as it slows
down, which is determined from $dE_e/dx$ for electrons and
$dE_\smI/dx$ for ions, can also be quite important, with larger ion
heating being more favorable for ignition.

Let us start by reviewing a study performed by N.M.~Hoffman and
C.L. Lee~\cite{hoff}, in which they compared the stopping power model
of `Corman-Spitzer'\,(CS)~\cite{cor} to that of C.K.~Li and
Petrasso~\cite{lip}. A significant difference between these two models
is that CS is valid only in a velocity window in
which the projectile speed is much greater than the thermal ion
velocity but much smaller than the thermal electron velocity, while Li
and Petrasso (LIP) is more general and was constructed to remain valid
for all non-relativistic velocities. The LIP stopping power tends to
be smaller than that of CS, and consequently the $\alpha$ particle
range for LIP is somewhat larger than for CS. By the aforementioned
considerations, we would therefore expect the ignition threshold for
LIP to be greater than that predicted from CS. Interestingly, however,
Hoffman and Lee found that both models produce the same ignition
threshold for the DT microspheres of Ref.~\cite{fraley74}, despite the
longer range of the $\alpha$ particle predicted by LIP~\cite{hoff}.
In addition to this, a preliminary study I have begun indicates that
other models, such as that of Ref.~\cite{leemore}, produce an ignition
threshold almost identical to that of CS and LIP.  As pointed out by
Ref.~\cite{hoff}, these observations can be explained by noting that
the models of CS and LIP differ most near the low energy thermal
regime of the $\alpha$ particle projectile, while their stopping
powers are almost identical at higher energies near the
\hbox{3.5\,MeV} threshold (at even higher energies, CS and LIP begin
to diverge, but this is well above the production threshold).
Consequently, the stopping power models of CS and LIP act to slow the
$\alpha$ particle down in almost exactly the same way throughout most
of its history, since they give equal values for $dE/dx$ throughout
most of the projectile energy regime, and this is why the ignition
profiles are virtually identical. In contrast, the stopping power of
LIP and BPS differ significantly even in the high energy regime
traversed by the $\alpha$ particle, and as we shall see, this produces
a marked difference in the ignition profiles of LIP and BPS.

Hoffman and Lee also observed that certain diagnostics, those sensitive
to small projectile energies of order the thermal background, are
quite different between CS and LIP.  For example, Hoffman and Lee
found LIP to be in agreement with experiment for the spectrum of fast
protons in a ${\rm D}\,^3{\rm He}$ filled capsule implosion, while CS
disagreed with the experimental \hbox{data~\cite{hicks00, hoff}}. The
diagnostic signatures explored by Hoffman and Lee tend to be more
sensitive to the low energy region of $dE/dx$, where CS and LIP have
the largest discrepancy, and this explains the differences in
diagnostics~\cite{hoff}. However, for diagnostics that are sensitive
to the soft regime of the projectile, one must compute $dE/dx$
extremely accurately. This presents a potential problem for models
that are inaccurate near the thermal regime, since their diagnostic
signatures are sensitive to this regime. However, as explained in
Section~II of Ref.~\cite{bps}, this criticism does not apply to the
BPS stopping power, and consequently BPS should provide quite reliable
diagnostics for such processes.

\section{\label{bpscontext} Context}

The basic physics responsible for charged particle stopping in a
plasma is energy exchange through Coulomb interactions, a process that
is quite similar to the Coulomb energy exchange that drives the
temperature equilibration between plasma components at differing
temperatures. Both of these processes, stopping power and temperature
equilibration, are local, in that they do not involve energy transport
in space. This should be contrasted with thermal conductivity, in
which heat flow by electrons or ions is not only driven by Coulomb
energy exchange, but also by the transport of energy from one spatial
location to another. In this sense, the basic physics of the stopping
power is identical to the physics of temperature equilibration between
plasma components, both of which differ somewhat from the physics of
thermal conductivity.  I will have more to say about thermal
conductivity in a future section, but for now, let us concentrate on
the rate at which the charged particle loses energy as it traverses
the plasma. For the Coulomb potential, a straightforward calculation
of the energy exchange rate diverges logarithmically in both the short
and long distance regimes~\cite{nuc}.  To obtain a finite result, we
must introduce {\em ad hoc} short and long distance cutoffs
$b_\text{min}$ and $b_\text{max}$ as we integrate over the impact
parameter $b$, in which case the rate takes the form
\begin{eqnarray}
  \frac{dE}{dt}
  = 
  K \int_{b_\text{min}}^{b_\text{max}} \frac{db}{b}
  = 
  K \hskip-0.6cm 
  \underbrace{\ln\!\left\{\frac{b_{\rm max}}{b_{\rm min}} 
  \right\}}_{\text{Coulomb Logarithm (CL)}}  
  \hskip-0.5cm .
\label{Krate}
\end{eqnarray}
The coefficient $K$ is an exactly calculable analytic prefactor that
depends critically on the physical process under consideration, while
the logarithmic term is called the Coulomb logarithm. The exact values
of the short and long distance cutoffs can only be estimated, and this
is a dominant source of non-systematic error in model
building. Choosing the values of $b_\text{min}$ and $b_\text{max}$
based on physical considerations (however well motivated they might
be), rather than direct calculations from theory, is what I will call
{\em model building}. We may increase the sophistication of models
such as Eq.~(\ref{Krate}) by including collective effects or improved
short distance physics, as in Ref.~\cite{lip}, but I will still refer
to this as {\em modeling}. In many cases in plasma physics, modeling
is essentially the only means by which to proceed, particularly in
complicated cases like warm dense matter. In some situations, however,
such as the weakly coupled plasma of a burning ICF capsule or the hot
plasma at the center of the sun, we can avail ourselves to simple
perturbative analytic techniques rather than model building, and this
is the path taken by BPS.

Before turning to the rigorous calculation of the stopping power, let
us investigate further the model building methodology and the physical
arguments used to choose the short and long distance cutoffs. Debye
screening sets the scale for the long distance cutoff, and we expect
on physical grounds that $b_{\rm max}=c_\smM\,\kappa^{-1}$ with the
dimensionless coefficient \hbox{$c_\smM \sim 1$}, where $\kappa$ is
the Debye wavenumber. The coefficient $c_\smM$ is
by no means a constant, but rather, it is typically a function of the
various plasma parameters (and the various projectile parameters in
the case of stopping power), which I will write schematically as
$c_\smM=c_\smM(m,T,n)$. Short of a rigorous calculation, there is no
prescription for determining the value or functional form of $c_\smM$,
and indeed, {\em choosing} the functional form of $c_\smM$ is part of
the model construction (usually $c_\smM$ is set unity).  There are
also various choices for the Debye wavenumber $\kappa$; for example,
should one use the electron screening length $\kappa_e^{-1}$ or the
total screening length $\kappa_\smD^{-1}$ determined from the
electrons and ions? The answer to this question is of course process
dependent; for example, if the ions can be treated as static and
screened by the electrons, then we should take
$b_\text{max}=c_\smM\,\kappa_e^{-1}$. However, to be rigorous, such
choices must come out of the calculation rather than being put into
the calculation by hand.  The situation for $b_\text{min}$ is even
less clear. The short distance cutoff is set by scattering. In the
extreme classical regime, we take $b_{\rm min}^\smC = c_m^\smC \times
(\text{distance of closest approach})$, and in the extreme quantum
regime we take $b_{\rm min}^\smQ = c_m^\smQ \times (\text{momentum
transfer})/\hbar$. In either case, the functions $c_m^\smC$ and
$c_m^\smQ$ are of order unity, but absent a rigorous calculation, this
is all we can really say about them.  To interpolate between the
extreme quantum and classical regimes, one usually employs an {\em ad
hoc} scheme such as $b_{ \rm min}=\left[\, (b_{\rm min}^\smC)^2 +
(b_{\rm min}^\smQ)^2\, \right]^{1/2}$, although in Ref.~\cite{bs} it
was shown that this simple interpolation is missing an important
logarithmic contribution. From these considerations, it is clear that
the model building process is at best logarithmically accurate, that
is to say, that the coefficient inside the Coulomb logarithm is known
only to within a factor of order unity, and a complete and rigorous
calculation is the only way to really settle the issue (this is why it
is not surprising that the coefficient inside the logarithm
varies by an order of magnitude within the literature, and with few
exceptions, there is no reason to prefer one model over
another
\footnote{\footnoteskip
  There are {\em calculations} in the literature, such as that of
  Gould and DeWitt~\cite{gd66}, that do obtain the correct
  coefficients inside the logarithm. However, to my knowledge, none of
  these calculations are systematic, in that they retain spurious
  higher-order terms and do not provide an error estimate.  Therefore,
  without independent verification such as BPS, these calculations do
  not have the power to know just how accurate they are.
}).  

Another point worth stressing is that the Coulomb logarithm itself is
process dependent, despite the fact that the same basic physics of
Coulomb energy exchange is at work in the stopping power and
temperature equilibration. For example, if we consider the charged
particle stopping power and the electron-ion temperature equilibration
rate, we can write
\begin{eqnarray}
  \frac{dE_p}{dx}
  &=& 
  K_p \, \ln\!\left\{\frac{b_{\rm max}}{b_{\rm min}} \right\} 
\label{dedxmodel}
\\[10pt] 
  \frac{d{\cal E}_{e \smI}}{dt} 
  &=& 
  K_{e\smI} \,\ln\!\left\{\frac{\bar b_{\rm max}}
  {\bar b_{\rm min}} \right\} \,\Big(T_e - T_\smI \Big) \ .
\label{dedtmodel}
\end{eqnarray}
The first expression (\ref{dedxmodel}) represents the energy loss per
unit distance of a projectile [the charged particle stopping power],
while the second expression (\ref{dedtmodel}) is the energy exchange
rate per unit volume between electrons at temperature $T_e$ and ions
at temperature $T_\smI$ [this rate is proportional to the temperature
difference, which I have explicitly indicated in
Eq.~(\ref{dedtmodel})].  I have placed a bar over the short and long
distance cutoffs in Eq.~(\ref{dedtmodel}) to indicate that the Coulomb
logarithm need not be the same as for other processes.  Concentrating
on the charged particle stopping power from here on, I will rewrite
the generic stopping power model (\ref{dedxmodel}) in the form
\begin{eqnarray}
  \frac{dE^\text{model}}{dx}
  &=& 
  K_\smCS \, \ln\!\Lambda_\text{model} \ .
\label{modeldef}
\end{eqnarray}
Here, I am emphasizing that the leading order coefficient $K_\smCS$ 
is exactly known~\cite{cor}, 
while the coefficient $\Lambda_\text{model}$ inside the logarithm is
only known to an order of magnitude.  The reason I do not call the BPS
calculation a model is because Ref.~\cite{bps} calculated the terms
under the logarithm exactly from first principles, including the
quantum to classical transition, along with a {\em controlled}
estimate of the error, which, in contrast to Eq.~(\ref{modeldef}), I
will write as
\begin{eqnarray}
  \frac{dE^\smBPS}{dx}
  &=& 
  K_\smCS \, \ln\!\Lambda_\smBPS + 
  \underbrace{~\text{controlled error term}~}_{\text{ small in weakly 
  coupled plasma}} \ .
\label{dEdxBPStoy}
\end{eqnarray}

As I will discuss more fully in the next section, any thermodynamic
quantity in a plasma can be expanded in {\em integer} powers of a
dimensionless plasma coupling parameter $g$~\cite{by}. Measuring
temperature in energy units and taking the electrostatic units to be
rationalized Gaussian, the choices I will employ throughout this
paper, the coupling takes the form
\footnote{\footnoteskip
  In ordinary nonrationalized Gaussian units we would write $g=e^2
  \kappa/T$. There is an independent coupling parameter for each
  plasma component, and this must be taken into account for a real
  calculation; however, for the purposes of explanation, we may
  consider an isolated plasma component. Finally, the usual plasma
  coupling $\Gamma$, defined in terms of the interparticle spacing,
  is related to the expansion parameter by $g^2 \propto \Gamma^3$.
  Integer expansions in $g$ are therefore expansions in fractional
  powers of $\Gamma$. 
}
\begin{eqnarray}
  g = \frac{e^2 \kappa}{4\pi}\,\frac{1}{T} \ ,
\label{gdef}
\end{eqnarray}
where $\kappa$ is the Debye wavenumber. The $g$-expansion admits
nonanalytic terms such as $\ln g$, and indeed, such terms are
essential in capturing the interplay of short and long distance
physics. As we shall see in the next section, the stopping power may
be systematically expanded in the form
\begin{eqnarray}
  \frac{dE^\smBPS}{dx} 
  =  
  -\underbrace{A\, g^2\ln g}_\text{LO}
  \,\,\,\, + \,\,
  \underbrace{~B g^2~}_\text{NLO} 
  \,\,+\,\,  
  {\cal O}(g^3) \ ,
\label{dedxNLO}
\end{eqnarray}
where I have indicated the leading order (LO) and the next-to-leading
order (NLO) terms. To get a feel for the size of $g$ in a plasma, at
the center of the sun we find $g=0.04$.  In a weakly coupled plasma in
which $g \ll 1$, the error terms, which I have denoted by ${\cal
O}(g^3)$, are quite small and the expansion (\ref{dedxNLO}) will be
near-exact, provided of course that we know the coefficients $A$ and
$B$ exactly.  The coefficient $A=A(m,T,n)$ is well known,
while $B=B(m,T,n)$ was calculated by
BPS~\cite{bps}, and as a matter of completeness, the final results of
the BPS calculation are displayed in Appendix~\ref{BPSdEdx}. To make
the connection with expression (\ref{dEdxBPStoy}), I will write the
leading order coefficient as $K_\smCS=A g^2$, and define the
dimensionless coefficient $C=\exp\{-B/A\}$. We can then express the
rate (\ref{dedxNLO}) in the form
\begin{eqnarray}
  \frac{dE^\smBPS}{dx} 
  &=& 
  K_\smCS\ln\Lambda_\smBPS \,+\, {\cal O}(g^3) \ ,
  ~~~\text{with}~~~
  \ln\Lambda_\smBPS = -\ln\left\{ C g\right\} \ .
\label{lngsqu}
\end{eqnarray}
This gives the {\em exact} Coulomb logarithm since we can calculate
$C=C(m,T,n)$. Incidentally, and this cannot be overemphasized, this
perturbative methodology {\em obviates} the need for the cutoff
parameters $b_\text{min}$ and $b_\text{max}$, and they can be
dispensed with from here out. The long and short distant cutoffs are
really only part of a heuristic device that, in my opinion, often
introduces more confusion than it purports to settle.

As discussed at length in Ref.~\cite{bps}, to leading and
next-to-leading order, and only to this order, can we decompose the
stopping power into its contribution to the various plasma species,
\begin{eqnarray}
  \frac{dE^\smBPS}{dx} 
  &=& 
  {\sum}_b   \frac{dE^\smBPS_b}{dx} 
  =
  {\sum}_b   K_b \ln\Lambda^\smBPS_b \,+\, {\cal O}(g^3) \ ,
\label{lngsqub}
\end{eqnarray}
where we identify $dE^\smBPS_b/dx$ as the stopping power contribution
{\em from} species~$b$. The notion of dividing the energy into contributions 
uniquely associated with individual plasma species is valid only
to order $g^2$, as three-body and higher collective effects render
this division meaningless when working to order $g^3$ and higher. For
the BPS stopping power, the quantity $dE^\smBPS_b/dx$ depends not just
upon parameters associated with the $b$-species, but upon the plasma
parameters of all other species through the dielectric function. This
should be contrasted with the corresponding quantity $dE_b^\smLIP/dx$
for LIP (except for the Coulomb logarithm~$\ln\Lambda_b^\smLIP$),
which depends only upon the plasma conditions of species $b$.
Since we shall be comparing BPS with LIP, I will close this section
by presenting the LIP stopping power in some detail.

Li and Petrasso~\cite{lip} modeled $dE/dx$ by
combining a generalization of the Fokker-Planck equation to account
for short-distance collisions and a well-chosen term involving a step
function to include long-distance collective effects.  They define a
Coulomb logarithm by using a minimum classical impact parameter that
interpolates between the classical and quantum regimes, as described
earlier. Using rationalized cgs units for the electric charge, the LIP
stopping power for species $b$ can be written
\begin{equation}
  \frac{d E^\smLIP}{dx} 
  =
  {\sum}_b \frac{d E^\smLIP_b}{dx}
  =
  {\sum}_b\frac{e_p^2}{4\pi}\,
  \frac{\kappa_b^2}{\beta_b m_b v_p^2} \left[G\left( \frac{1}{2}\,
  \beta_b m_b v_p^2 \right) \, \ln \Lambda^\smLIP_b + H \left( \frac{1}{2}\,
  \beta_b m_b v_p^2 \right) \right] \, ,
\label{lii}
\end{equation}
where the contribution to species $b$ is $d E^\smLIP_b/dx$.
Here, the function multiplying the Coulomb logarithm is defined by
\begin{equation}
  G(y) = \left[ 1 - \frac{ m_b}{m_p}\,
  \frac{d}{dy} \right] \mu(y) \,,
\label{G}
\end{equation}
where 
\begin{equation}
  \mu(y) = \frac{2}{\sqrt\pi} \int_0^y dz \,
  z^{1/2} \, e^{-z} \,,
\end{equation}
and
\begin{equation}
  H(y) = \frac{m_b}{m_p} \left[ 1 + \frac{d}{dy}
  \right] \mu(y) + \theta(y-1) \, \ln \left( 2
  e^{-\gamma} y^{1/2} \right) \,,
\label{H}
\end{equation}
with $\theta(x)$ being the unit step function: $\theta(x) = 0$ for
$x<0$ and $\theta(x) = 1$ for $x>0$. 
Li and Petrasso define a Coulomb logarithm in terms of the combination
of classical and quantum cutoffs as described above, namely
\begin{equation}
  \ln \Lambda_b^\smLIP = - \frac{1}{2} \ln \kappa_\smD^2 \,
  B_b^2 \,,
\end{equation}
where the short distance cutoff is defined by 
\begin{equation}
  B_b^2 = \left( \frac{\hbar}{2 m_{pb} u_b }
  \right)^2 + \left( \frac{ e_p e_b}{4\pi m_{pb} u_b^2 }
  \right)^2 \,,
\end{equation}
in which $m_{pb}=m_p m_b/(m_p+m_b)$ is the reduced mass of the
projectile and species $b$, and 
\begin{equation}
  u_b^2 = v_p^2 + \frac{2}{\beta_b m_b }
\end{equation}
defines an average of the squared projectile and thermal velocities.
In rationalized units, the Debye wave number is
\begin{eqnarray}
  \kappa_b^2 &=& \frac{e_b^2\, n_b}{T_b} \ .
\label{kbdef}
\end{eqnarray}
In future sections, we will also need the plasma frequency, 
and in rationalized units it takes the form 
\begin{eqnarray}
  \omega_b^2 &=& \frac{e_b^2\, n_b}{m_b} \ .
\label{wbdef}
\end{eqnarray}

\section{Analytic Considerations: The BPS Calculation}

In this section I will review the salient features of the BPS
calculation~\cite{bps}. This calculation includes both hard (short
distance) physics and {\em dynamic} collective (long distance)
physics, joined together exactly and unambiguously, and systematized
by a power series expansion in the plasma coupling constant $g$.
Ref.~\cite{bps} calculates both the charged particle stopping power
and the electron-ion temperature equilibration rate, thereby providing
an exact calculation of the coefficient inside the Coulomb logarithm
for both processes in (\ref{dedxmodel}) and (\ref{dedtmodel}). 
An additional feature of the BPS calculation is
that it also provides an exact interpolation between the extreme
classical and extreme quantum regimes.  The calculation exploits a
procedure in quantum field theory known as dimensional
regularization, or dimensional continuation as I will call it
here. The basis idea is that divergent theories exhibit finite poles
of the form $1/(\nu-3)$ when analyzed in an arbitrary number of
spatial dimensions $\nu$. We can then manipulate finite quantities in
such a way that we preserve the delicate relation between long and
short distance physics. In a physical process, the divergent pole
terms cancel and we can set the number of spatial dimensions to
$\nu=3$ at the end of the calculation, thereby giving a finite result
with correct long and short distance physics.

To start the discussion, let ${\bf x}$ and ${\bf v}$ denote the
\hbox{$\nu$-dimensional} position and velocity vectors of a particle.
The Coulomb potential for two particles separated a distance $r=\vert
{\bf x} -{\bf x}^\prime\vert $ is $V_\nu(r)=C_\nu\, e^2/r^{\nu-2}$,
where $C_\nu=\Gamma(\nu/2-1)/ 4\pi^{\nu/2}$ is a spatially dependent
geometric factor. This potential follows directly from a simple
multidimensional generalization of Gauss' Law.  For every species $b$,
the single-particle distribution function $f_b$ will be defined so
that $f_b({\bf x},{\bf v},t)\, d^\nu x \,d^\nu v$ gives the number of
particles of species $b$ in a small hypervolume $d^\nu x$ about ${\bf
x}$ and $d^\nu v$ about ${\bf v}$ at time $t$. We shall take the
plasma components to be Maxwell-Boltzmann, although with more work,
the situation can be generalized to Fermi-Dirac statistics as
well~\cite{bs}. In the case of a projectile $p$, the distribution
function $f_p$ will be peaked about a specific point in phase space,
and the stopping power can then be calculated by
\begin{eqnarray}
  \frac{dE_p}{dx}
  =
  \frac{1}{v_p}\,\frac{dE_p}{dt}
  =
  \frac{1}{v_p}\!\int\! d^\nu x \,d^\nu p~ \frac{p^2}{2m_p}\,
  \frac{\partial f_p}{\partial t} \ .
\label{dEdtdef}
\end{eqnarray}
Expression (\ref{dEdtdef}) is, of course, extremely problematic in
three dimensions, but completely finite when $\nu \ne 3$. We can gain
some additional insight into the divergence problem, however, by
returning to three dimensions for a moment. In three dimensions, if we
knew the {\em exact} form of $f_p$ in the background of the other
distribution functions $f_b$ of the various plasma components, then
Eq.~(\ref{dEdtdef}) would be completely finite and well defined in
three dimensions.  Unfortunately, the requisite solution for $f_p$ can
only be obtained by solving the full BBGKY hierarchy of kinetic
equations, an impossible feat. The divergence problem in
Eq.~(\ref{dEdtdef}) arises only because we must use an approximation
for $f_p$, usually obtained by truncating the BBGKY equations to a
first-order kinetic equation, such as the Boltzmann or Lenard-Balescu
equation, and these approximates miss correlations that would
otherwise render Eq.~(\ref{dEdtdef}) finite in three spatial
dimensions. Curiously, these divergences occur only for the Coulomb
potential, and only then in three spatial dimensions!

These observations provide a path forward. Let us return to an
arbitrary number of dimensions, where the Coulomb potential provides
finite results. Let us also define multipoint correlation functions in
a similar manner to the $f$'s, and in this way we can construct the
full BBGKY hierarchy in an arbitrary number of dimensions. For
simplicity I will drop the subscript on the distribution functions.
In dimensions $\nu>3$, the standard textbook derivation of the
Boltzmann Equation (BE) goes through without the infrared divergent
scattering kernel found in three dimensions, and I will denote this
finite \hbox{$\nu$-dimensional} scattering kernel by the shorthand
notation $B_\nu[f]$. The important point to emphasize here is that the
\hbox{$\nu$-dimensional} Coulomb potential $V_\nu \sim e^2/r^{\nu-2}$
emphasizes short-distance over long-distance physics when $\nu>3$, and
this means that the BBGKY hierarchy reduces to the Boltzmann equation
to {\em leading} order in $g$ in these dimensions:
\begin{eqnarray}
  {\rm BBGKY} \Rightarrow
  \frac{\partial f}{\partial t} + 
  {\bf v} \! \cdot \!{\bm\nabla}_{\!x}\,  f 
  =
  B_\nu[f] 
  ~~~{\rm to~LO~in~} g {\rm ~for~}\nu > 3\ .
\label{BEsimpnu}
\end{eqnarray}
Here, the $\nu$-dimensional spatial gradient has been denoted by
${\bm\nabla}_{\!x}$.  In dimensions $\nu<3$, the standard textbook
derivation of the Lenard-Balescu equation (LBE) goes through without
the ultraviolet divergent scattering kernel found in three dimensions,
and I will denote this finite \hbox{$\nu$-dimensional} scattering
kernel by the shorthand notation $L_\nu[f]$.  In dimensions $\nu<3$,
the Coulomb potential $V_\nu(r)$ emphasizes long-distance physics over
short-distance effects, and consequently, to leading order in $g$, the
BBGKY hierarchy reduces to the Lenard-Balescu equation (LBE) in this
spacial regime:
\begin{eqnarray}
  {\rm BBGKY} \Rightarrow
  \frac{\partial f}{\partial t} + 
  {\bf v} \! \cdot \!{\bm\nabla}_{\!x}\, f
  =
  L_\nu[f]
  ~~~{\rm to~LO~in~} g {\rm ~for~}\nu < 3 \ .
\label{LBEsimpnu}
\end{eqnarray}
Space does not permit us to write down the exact forms of $B_\nu[f]$
and $L_\nu[f]$ here, but one may consult Ref.~\cite{bps} for these
expressions. These kinetic equations allow one to calculate the
stopping power in $\nu>3$ and $\nu<3$, the results of which are
presented in Sections~8 and 7 of Ref.~\cite{bps}, respectively. These
calculations involve performing a series of momentum and wave number
integrals in arbitrary dimensions $\nu$, and they take form~\cite{bps}
\begin{eqnarray}
  \frac{dE_p^\smGT}{dx}
  &=& 
  \frac{1}{v_p}\int\! d^\nu p\, \frac{p^2}{2m_p}\,B_\nu[f]
  =
  H(\nu)\,\frac{g^2}{\nu-3} 
  +
  {\cal O}(\nu-3) 
  \hskip0.66cm:~  {\rm LO~in}~g~{\rm when~}\nu > 3 \ ,
\label{dedtonecal}
\\[5pt]
  \frac{dE_p^\smLT}{dx}
  &=&
  \frac{1}{v_p}\int\! d^\nu p\, \frac{p^2}{2m_p}\, L_\nu[f]
  =
  G(\nu)\, \frac{g^{\nu-1}}{3-\nu} 
  + {\cal O}(3-\nu) 
  \hskip0.7cm :~ {\rm LO~in}~g~{\rm when~} \nu < 3 \ .
\label{dedttwocal}
\end{eqnarray}
The analytic expressions for $H(\nu)$ and $G(\nu)$ are rather
complicated
,\footnote{\footnoteskip
  For the related process of electron-ion temperature equilibration,
  in contrast, the expressions for $H(\nu)$ and $G(\nu)$ are quite
  simple. 
}
and space does not permit their reproduction here. In this paper, we
are only interested in their analytic properties as a function of
$\nu$. In particular, the coefficients $H(\nu)$ and $G(\nu)$ can be
expanded in powers of $\epsilon=\nu-3$, and we find
\begin{eqnarray}
  H(\nu) 
  =
  -A + \epsilon \,H_1 + {\cal O}(\epsilon^2)
  ~~~{\rm and}~~~
  G(\nu) 
  =
  -A + \epsilon \,G_1 + {\cal O}(\epsilon^2) \ .
\label{Gexp}
\end{eqnarray}
For our purposes, we do not require the exact forms of $H_1$, $G_1$,
nor that of the leading term~$A$. It is sufficient to note that the
leading terms in Eq.~(\ref{Gexp}) are equal, so that \hbox{$H(\nu
\equiv 3)=G(\nu \equiv 3)$}.  This is a fact that arises from the
calculation itself, as it must, and it should be emphasized that this
equality is not arbitrarily imposed by hand.  It is a {\em crucial}
point that the leading terms are identical, as this will allow the
short- and long-distance poles to cancel, thereby giving a finite
result.

Since the rates $dE_p^\smGT/dx$ of Eq.~(\ref{dedtonecal}) and
$dE_p^\smLT/dx$ of Eq.~(\ref{dedttwocal}) were calculated in mutually
exclusive dimensional regimes, one might think that they cannot be
compared. However, even though Eq.~(\ref{dedttwocal}) was originally
calculated in $\nu<3$ for integer values of $\nu$, we can analytically
continue the quantity $dE_p^\smLT/dx$ to values $\nu>3$ (for
simplicity I have been omitting the functional dependence on $\nu$
from the stopping power, but it is there nonetheless).  This process
of analytic continuation to non-integer values of dimension is analogous
to way in which the factorial function $n!$ on the positive integers
can be generalized to the Gamma function $\Gamma(z)$ over the complex
plane, including both the positive and negative real axes.  We can
then directly compare Eqs.~(\ref{dedtonecal}) and
(\ref{dedttwocal}). Upon writing the $g$-dependence of
Eq.~(\ref{dedttwocal}) as $g^{2 + (\nu-3)}$, when $\nu>3$ we see that
Eq.~(\ref{dedttwocal}) is indeed higher order in $g$ than
Eq.~(\ref{dedtonecal}):
\begin{eqnarray}
  \frac{dE_p^\smLT}{dt}
  &=&
  -G(\nu)\, \frac{g^{2+(\nu-3)}}{\nu-3} 
  +
  {\cal O}(\nu-3) 
  ~~:~{\rm NLO~in}~g~{\rm when~} \nu > 3 \ .
\label{NLOgterm}
\end{eqnarray}
By power counting arguments, no powers of $g$ between $g^2$ and
$g^{\nu-1}$ can occur in Eq.~(\ref{dedtonecal}) for $\nu>3$, and
therefore Eq.~(\ref{dedttwocal}) indeed provides the correct
next-to-leading order term in $g$ when the dimension is analytically
continued to $\nu>3$. The individual pole-terms in
Eqs.~(\ref{dedtonecal}) and (\ref{NLOgterm}) will cancel giving a
finite result when the leading and next-to-leading order terms are
added. The resulting finite quantity will therefore be accurate to
leading and next-to-leading order in $g$ as the $\nu \to 3$ limit is
taken:
\begin{eqnarray}
  \frac{dE_p}{dx}
  =
  \lim_{\nu \to 3^+}
  \Bigg[
  \underbrace{~\frac{dE^\smGT}{dx}~}_{\rm LO}
  +
  \underbrace{~\frac{dE^\smLT}{dx}~}_{\rm NLO}
  \Bigg] + {\cal O}(g^3) \ .
\label{dedtorderg3}
\end{eqnarray}
Note that this does not lead to any form of ``double counting'' since
we are merely adding the next-to-leading order term (\ref{NLOgterm})
to the leading order term (\ref{dedtonecal}) at a common value
of~\hbox{$ \nu > 3$}.  We are now in a position to evaluate the limit
in Eq.~(\ref{dedtorderg3}). Defining $\epsilon=\nu-3$ as before, note
that $g^{\epsilon} = \exp\{\epsilon \ln g\} = 1 + \epsilon\ln g +
{\cal O}(\epsilon^2)$, which gives the relation
\begin{eqnarray}
  \frac{g^\epsilon}{\epsilon}
  = 
  \frac{1}{\epsilon} 
  +
  \ln g + {\cal O}(\epsilon) \ .
\label{gexp}
\end{eqnarray}
Substituting Eq.~(\ref{gexp}) into Eq.~(\ref{NLOgterm}), adding this
result to Eq.~(\ref{dedtonecal}), and then taking the limit gives
\begin{eqnarray}
  \frac{dE_p}{dx}
  =
  - A\, g^2 \ln g   +  B\, g^2  + {\cal O}(g^3) \ ,
\end{eqnarray}
with $B=H_1-G_1$, in agreement with Eq.~(\ref{dedxNLO}).  In this way,
BPS has calculated the charged particle stopping power accurate to
leading order and next-to-leading order in $g$. For completeness, the
full BPS stopping power is presented in Appendix~\ref{BPSdEdx}.

\newcommand{\RaBPS}{30\mu\text{m}}
\newcommand{\EiBPS}{0.38\,{\rm MeV}}
\newcommand{\EeBPS}{3.16\,{\rm MeV}}
\newcommand{\RaLIP}{25 \mu\text{m}}
\newcommand{\EiLIP}{0.45\,{\rm MeV}}
\newcommand{\EeLIP}{3.09\,{\rm MeV}}
\newcommand{\DRa}{+20\%}
\newcommand{\DEi}{-16\%}
\newcommand{\DEe}{+2\%}
\begin{figure}
\includegraphics[scale=0.45]{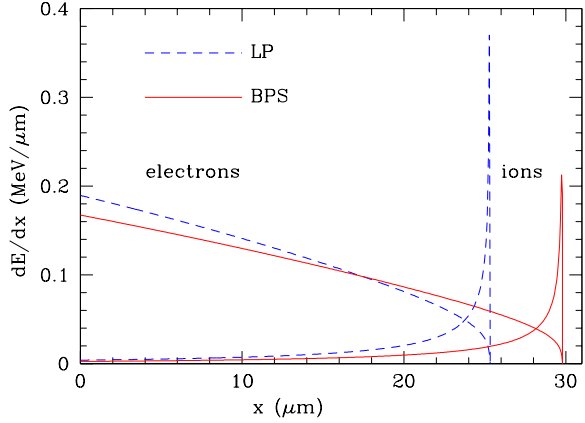}
\vskip-0.7cm 
\caption{\footnoteskip 
  A comparison of the stopping power of BPS to
  Li and Petrasso (LIP) for a 3.5\, MeV $\alpha$ particle in a
  equimolar DT plasma. The plasma temperature is $T=3\,{\rm keV}$ and
  the electron number density is $n_e=1 \times 10^{25}\,{\rm
  cm}^{-3}$, giving a coupling $g=0.01$.  For BPS, the range and the
  energy deposited into ions and electrons are 
  $R_\alpha^\smBPS=\RaBPS$, 
  $E_\smI^\smBPS=\EiBPS$, and
  $E_\text{e}^\smBPS=\EeBPS$, respectively.  For LIP, these
  corresponding quantities are 
  $R_\alpha^\smLIP=\RaLIP$,
  $E_\smI^\smLIP=\EiLIP$, and 
  $E_\text{e}^\smLIP=\EeLIP$. 
  The percent differences between BPS and LIP are: 
  $\Delta R_\alpha=\DRa$, 
  $\Delta E_\smI=\DEi$, and 
  $\Delta E_\text{e}=\DEe$,
  where $\Delta X \equiv (X^\smBPS-X^\smLIP)/X^\smLIP$.  }
\label{fig:BPSvsLIPa}
\end{figure}
\newcommand{\RaBPSb}{3.7 \mu\text{m}}
\newcommand{\EiBPSb}{2.00\,{\rm MeV}}
\newcommand{\EeBPSb}{1.51\,{\rm MeV}}
\newcommand{\RaLIPb}{2.9 \mu\text{m}}
\newcommand{\EiLIPb}{-}
\newcommand{\EeLIPb}{-}
\newcommand{\DRab}{-}
\newcommand{\DEib}{-}
\newcommand{\DEeb}{-}
\begin{figure}
\includegraphics[scale=0.45]{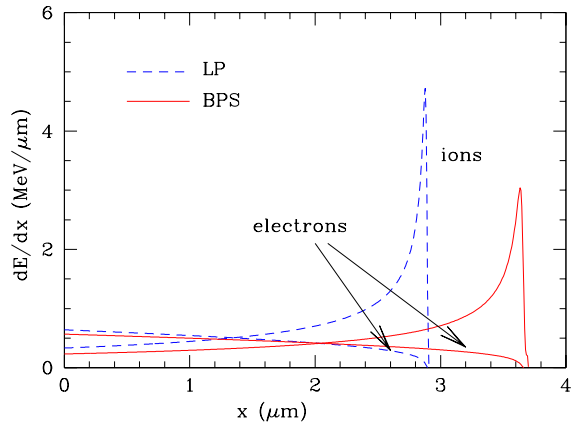}
\vskip-0.7cm 
\caption{\footnoteskip 
  As in Fig.~\ref{fig:BPSvsLIPa}, BPS {\em
  vs}. LIP for a 3.5\,MeV $\alpha$ particle, in a DT plasma at
  $T=30\,{\rm keV}$ and $n_e=1 \times 10^{27}\,{\rm cm}^{-3}$, giving
  a plasma coupling is $g=0.003$; 
  $R_\alpha^\smBPS=\RaBPSb$, 
  $E_\smI^\smBPS=\EeBPSb$, and 
  $E_\text{e}^\smBPS=\EeBPSb$; 
  $R_\alpha^\smLIP=\RaLIPb$, 
  $E_\smI^\smLIP=2.15\,{\rm MeV}$, and 
  $E_\text{e}^\smLIP=1.36\,{\rm MeV}$; 
  $\Delta R_\alpha=28\%$, 
  $\Delta E_\smI=7\%$, and 
  $\Delta E_\text{e}=11\%$. 
}
\label{fig:BPSvsLIPb}
\end{figure}

As alluded to earlier, the BPS calculation predicts the range of the
3.5 MeV $\alpha$ particle in a hot DT plasma to be 20--30\% longer
than typical plasma models in the literature with a smaller energy
deposition into the ions. For LIP and BPS, this is illustrated
Figs.~\ref{fig:BPSvsLIPa} and \ref{fig:BPSvsLIPb}. As we shall see in
the next section, the longer $\alpha$ particle range and less
efficient ion heating tend to make ignition more difficult to
achieve.

\section{Numerical Considerations: Effects of Stopping Power on Ignition}


Since this is a preliminary study, which I would like to keep as clean
and simple as possible, I will not model a real ICF capsule
here. Instead, I would like to look at the essential features of the
stopping power without the complications of additional processes like
hydrodynamic instabilities and thermal conductivity. I have already
mentioned the latter problem, but I would like to make a few more
comments. In a real ICF capsule, consistency requires that we model or
calculate the thermal conductivity just as accurately in $g$ as the stopping
power. However, calculating the heat flow is harder than calculating
the stopping power: one must not only contend with the short and long
distance physics of the instantaneous Coulomb interactions, but one
must also consider the rate of energy transport from one spatial
location to another.  It is not clear to me whether the Coulomb
logarithm of the thermal conductivity can be calculated with the BPS
methodology, since one must invert a heat kernel in addition to
performing the appropriate multidimensional integrals. One must also
consider the possibility of $\alpha$ particles emerging from the hot
spot into the surrounding colder and more strongly coupled plasma, a
region where the BPS stopping power may not apply. Including these
effects is of course essential, but at this stage, they are somewhat
premature.


\begin{figure}
\includegraphics[scale=0.50]{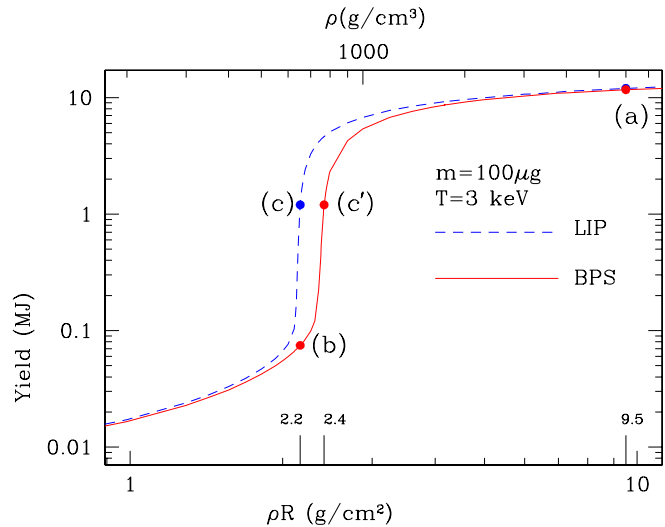}
\vskip-0.7cm 
\caption{\footnoteskip 
The yield in MJ as a function of initial $\rho
R$ in ${\rm g/cm^2}$ for a DT microsphere of Ref.~\cite{fraley74}.
The initial uniform temperature is taken to be $T_0=3\,{\rm keV}$ and
the initial mass of the sphere is $m_0=100\,{\rm \mu g}$.  The upper
axis gives the initial density $\rho$ in ${\rm g/ cm^3}$. The dashed
curve is the yield profile for the stopping power of Li and Petrasso
(LIP), while the solid curve corresponds to that of Brown, Preston,
and Singleton (BPS). The points labeled by (a) and (b) correspond to
full yield 1\% maximum yield, respectively. The points labeled by (c)
and (c$^\prime$) correspond to 10\% maximum yield, the first for LIP
and the second for BPS.  The BPS stopping power has a longer $\alpha$
particle range than LIP, and it delivers less $\alpha$ particle energy
to the ions. This has the effect of increasing the ignition threshold
for BPS relative to LIP.  }
\label{fig:ignitioncurve}
\end{figure}

For the purposes of this study, it suffices to concentrate only on the
time period after maximum compression and minimum volume, near the
onset of thermonuclear burn. Stopping power plays less of a role in
the implosion process, which is set in motion by an assembly of high
intensity lasers or pulsed-power diodes that generate a radiation
field that ablates a material shell (such as beryllium or plastic)
encapsulating the DT gas. In fact, I will ignore the presence of a
surrounding shell completely, and instead look at the idealized case
of Fraley {\it et al.}~\cite{fraley74}. These authors consider an
initially static microsphere of compressed DT at uniform temperature
$T_0$ and density $\rho_0$.  Because of the non-zero pressure in the
microsphere, the outermost surface of the spherical assembly acquires
an outward velocity (comparable to the speed of sound in the DT
plasma), which produces an incoming rarefaction wave. The net effect
is that the center of the sphere remains slightly hotter than the
surface, and consequently thermonuclear burn (if it occurs) will start
at the center of the sphere and burn outward. The $\alpha$-particle
from the fusion reaction $D + T \to \alpha + n$ will provide the
bootstrap heating that will initiate ignition and maintain
thermonuclear burn.

For definiteness, we will take the initial temperature of these so
called Fraley spheres to be $T_0 = 3\,{\rm keV}$, with an initial mass
$m_0=100\,\mu{\rm g}$. We shall vary the initial density $\rho_0$ and
measure the final yield $Y$ produced in the simulation. To find the
final state of the system, we numerically integrate the Navier-Stokes equations
using a two-dimensional Lagrangian finite-difference scheme. Radiative
energy exchange is included via a multi-frequency diffusion
approximation.  Other transport processes are included as well,
and plasma electrons and ions are each assumed to be
characterized by separate Maxwellian thermal distributions, and
exchange energy at a rate characterized by Spitzer
theory
.\footnote{\footnoteskip
The BPS electron-ion 
coupling rate calculated in Ref.~\cite{bps} was also used, and
little difference between this near-exact result and Spitzer.
}
The results of these simulations are illustrated in
Fig.~\ref{fig:ignitioncurve}, where we plot the yield in MJ as a
function of initial $\rho R$ in ${\rm g/cm^2}$. The dashed curve is
the yield profile predicted by the LIP stopping power, while the solid
curve corresponds to that of BPS. The points labeled by (a) and (b)
correspond to full yield and 1\% of full yield, respectively. When
full burn is assured, as in case (a), note that the yield is
relatively independent of the stopping power.  The points labeled by
(c) and (c$^\prime$) correspond to the 10\% maximum yield threshold
for LIP and BPS respectively. As we see, the effect of BPS relative to
LIP is to push the 10\% threshold from $(\rho R)_\smLIP=2.2 \,{\rm
g/cm^2}$ to $(\rho R)_\smBPS=2.4 \,{\rm g/cm^2}$, an increase in $\rho
R$ of approximately $10\%$.

\section{Electron-Ion Temperature Equilibration}

The final question I shall address is the effect of electron-ion
temperature equilibration on the ignition threshold.  The rate at
which electrons and ions come into equilibrium,
\begin{eqnarray}
  \frac{{\cal E}_{e\smI}}{dt}
  =
  -\,{\cal C}_{e\smI}\,\big(T_e - T_\smI \big) \ ,
\end{eqnarray}
was also calculated in Ref.~\cite{bps} to the same level of rigor as
the stopping power, and this is presented in Appendix~\ref{BPStei}
in full generality. In the high temperature limit that applies to
ignition, the rate coefficient takes the simple form
\begin{eqnarray}
  {\cal C}_{e\smI}
  = 
  \frac{\omega_\smI^2}{2\pi}\, \kappa_e^2\,
  \sqrt{\frac{m_e}{2\pi\, T_e}}\, \ln\Lambda_\smBPS \ ,
  ~~~\text{with}~~~
  \ln\Lambda_\smBPS
  =
  \frac{1}{2}\left[\ln\!\left\{\frac{8 T_e^2}{\hbar^2 \omega_e^2}
  \right\} - \gamma - 1 \right] \ ,
\label{bpsrate}
\end{eqnarray}
where $\gamma=0.57721 \cdots$ is the Euler constant, $\kappa_e$ and
$\omega_e$ are the electron Debye wave number and plasma frequency,
and \hbox{$\omega_\smI^2 =\sum_i \omega_i^2$} is sum of the squares of
the ion plasma frequencies
.\footnote{\footnoteskip
  Equation~(\ref{bpsrate}) corresponds to
  Eqs.~(3.61)~and~(12.12) of Ref.~\cite{bps}, where I have taken this
  opportunity to correct a small transcription error: when passing from
  Eq.~(12.43) to Eq.~(12.44) in Ref.~\cite{bps}, a factor of 1/2 was
  dropped. Restoring this factor of 1/2 changes the additive constant
  outside the logarithm from the $-\gamma-2$ that appears in Eq.~(12.12)
  of Ref.~\cite{bps} to the constant $-\gamma-1$ in
  Eq.~(\ref{bpsrate}) of this paper.
} 
Comparing Eqs.~(\ref{bpsrate}) and (\ref{dedtmodel}), we again see
that there is no need to reference the heuristic scales $b_{\rm min}$
and $b_{\rm max}$. The BPS calculation predicts a smaller rate than
typical model calculations in the literature.  This is ostensibly
advantageous for ignition, since the ion temperature is more weakly
coupled to the electron temperature and can run away more easily;
however, initial simulations indicate that the electron-ion
equilibration has little effect upon ICF yield. Nonetheless, I would
expect this rate to have some effect upon burn diagnostics, and I
will explore this possibility in the future. 

\vskip-0.5cm 
\vbox{
\begin{acknowledgements}
\baselineskip 16pt
\vskip-0.5cm 
 I would like to thank Nelson Hoffman for many useful discussions and
 for his help on the numerical section of this work.
\end{acknowledgements}
}

\appendix
\section{\label{BPSdEdx} The BPS Charged Particle Stopping Power}

As a matter of completeness, I will present the full form of the BPS
stopping power calculated in Ref.~\cite{bps}.  Consider a projectile
of mass $m_p$ and charge $e_p$ moving through a multi component plasma
at speed $v_p$. Each plasma component is assumed to be in equilibrium
with itself but not necessarily with the other components, and will be
labeled by an index $b$ (including the electron component). The mass
of a component is $m_b$, the number density is $n_b$, and the
temperature is $T_b$. Temperature will be measured in energy units,
and I will use the notation $\beta_b=1/T_b$ for the inverse
temperature. To second order in the plasma coupling $g$, the order to
which we are working, the stopping power $dE/dx$ of the projectile can
meaningfully be broken into the contributions $dE_b/dx$ from the
separate components, so that $dE/dx = {\sum}_b dE_b/dx$.  Each
contribution can be expressed as a sum of three terms,
\begin{eqnarray}
  \frac{dE_b}{dx}
  =
  \left(\frac{dE^\smC_{b,\smS}}{dx}
  +
  \frac{dE^\smC_{b,\smR} }{dx} \right)
  +
  \frac{dE^\smDelQ_b}{dx} \ ,
\label{dEpdx}
\end{eqnarray}
where the first two arise from classical short and long distance
physics, and the latter term from short distance two-body quantum
diffraction. The stopping power $dE_b/dx$ is a function of the
parameters of the projectile and the plasma background.  For
convenience, however, we shall emphasize the functional dependence
only upon the projectile energy $E_p = \frac{1}{2} m_p v_p^2$. The
terms in Eq.~(\ref{dEpdx}) are given by Eqs.~(3.4), (3.3) and (3.19),
respectively, in BPS~\cite{bps}:
\begin{eqnarray}
\nonumber 
  \frac{dE^\smC_{b,\smS}}{dx}(E_p)
  &=&
  \frac{e_p^2}{4\pi}\,\frac{\kappa_b^2}{m_p v_p}\,
  \left(\frac{m_b}{2\pi \beta_b}\right)^{1/2}\,
  \int_0^1 \! du\, u^{1/2} \exp\!\Big\{-\frac{1}{2}\, 
  \beta_b m_b v_p^2\, u \Big\} 
\\&& 
  \Bigg[ 
  \Big(-\ln\!\Bigg\{\beta_b\,\frac{e_p e_b\,K}{4\pi}  \,
  \frac{m_b}{m_{pb}}\, \frac{u}{1-u} 
  \Bigg\} 
  + 2 - 2 \gamma \Big) \Big(\beta_b M_{pb} v_p^2 \Big) 
  + \frac{2}{u} \Bigg]
\label{dedxS}
\\[10pt]
\nonumber 
  \frac{dE^\smC_{b,\smR} }{dx}(E_p)
  &=&
  \frac{e_p^2}{2\pi}\,\frac{i}{2\pi}\int_{-1}^1 \! d\cos\theta\,
  \cos\theta\, \frac{\rho_b(v_p\cos\theta)}{\rho_\text{total}
  (v_p \cos\theta)}\,
  F(v_p\cos\theta) \ln\!\left\{\frac{F(v_p\cos\theta)}{K^2} \right\}
  +
\\[5pt]&&
  \frac{e_p^2}{2\pi}\,\frac{i}{2\pi}\,\frac{\rho_b(v_p)}
  {\rho_\text{total}(v_p)}\,\Big[
  F(v_p) \ln\!\left\{\frac{F(v_p)}{K^2}\right\} -
  F^*(v_p) \ln\!\left\{\frac{F^*(v_p)}{K^2} \right\}
  \Big]
\label{dedxR}
\\[10pt]
\nonumber 
  \frac{dE^\smDelQ_{b}}{dx}(E_p)
  &=&
  \frac{e_p^2}{4\pi}\,\frac{\kappa_b^2}{2 \beta_b m_p v_p^2}
  \int_0^\infty \! dv_{pb}\, \Big[2 \psi(1 + i \eta_{pb}) - 
  \ln \eta_{pb}^2\Big]
\\&&
\nonumber 
  \Bigg[ \Big[1 + \frac{M_{pb}}{m_b}\,\frac{v_p}{v_{pb}}\,
  \left(\frac{1}{\beta_b m_b v_p v_{pb}}-1\right) \Big] \Bigg. 
  \exp\!\left\{-\frac{1}{2}\,\beta_b m_b \left(v_p - v_{pb} \right)^2  
  \right\} +
\\&&
  \phantom{\Bigg]}
  \Big[1 + \frac{M_{pb}}{m_b}\,\frac{v_p}{v_{pb}}\,
  \left(\frac{1}{\beta_b m_b v_p v_{pb}}+1\right)  \Big]
  \exp\!\left\{-\frac{1}{2}\,\beta_b m_b \left(v_p + v_{pb} \right)^2  
  \right\} \Bigg] \ .
\label{dedxQ}
\end{eqnarray}
In the long distance contribution (\ref{dedxR}), we define $\theta$ as
the angle between the vectors ${\bf k}$ and ${\bf v}_p$, while $F$ is
equivalent to the plasma dielectric function,
\begin{eqnarray}
  k^2 \epsilon(k,\omega={\bf k}\cdot{\bf v}_p) = k^2 + 
  F(v_p \cos\theta)
\hskip0.5cm \text{in which}\hskip0.5cm 
  F(v) 
  = 
  \int_{-\infty}^\infty \! du \, 
  \frac{\rho_\text{total}(u)}{v - u + i\eta} \ ,
\label{Fdef}
\end{eqnarray}
with 
\begin{eqnarray}
  \rho_\text{total}(v)
  =
  {\sum}_b\, \rho_b\!\left(v\right) 
\hskip0.5cm \text{and} \hskip0.5cm 
  \rho_b(v) 
  = 
  \kappa_b^2\,\sqrt{\frac{\beta_b m_b}{2\pi}}\, v\,
  \exp\!\left\{-\frac{1}{2}\,\beta_b m_b\, v^2\right\} \ .
\label{rhototdefA}
\end{eqnarray}
In Eq.~(\ref{dedxQ}) the integration variable is 
\begin{eqnarray}
  v_{pb}=\vert {\bf v}_p - {\bf v}_b \vert
\label{Defvpb}
\end{eqnarray}
and the dimensionless quantum parameter is 
\begin{eqnarray}
  \eta_{pb}  = \frac{e_p e_b}{4\pi \hbar v_{pb}} \ .
\label{etapb}
\end{eqnarray}
The Diagamma function is defined by $\psi = \Gamma^{-1}d\Gamma/dz$,
so that
\begin{eqnarray}
  \text{Re}\,\psi(1 + i \eta) = \sum_{k=1}^\infty \frac{1}{k}\,
  \frac{\eta^2}{k^2+\eta^2} - \gamma \ ,
\label{Repsi}
\end{eqnarray}
where $\gamma=0.5572 \cdots$ is the Euler constant. 
The total mass and the reduced mass are 
\begin{eqnarray}
  M_{pb} 
  =
  m_p + m_b
\hskip1cm 
  \frac{1}{m_{pb}}
  =
  \frac{1}{m_p} + \frac{1}{m_b} \ .
\label{Mmab}
\end{eqnarray}
The sum of terms (\ref{dedxR}) and (\ref{dedxS}) form the classical
contribution, and the factor $K$ is an arbitrary wave number that
cancels in the sum of Eqs.~(\ref{dedxR}) and (\ref{dedxS}). It is
convenient to set $K=\kappa_e$.

\section{\label{BPStei} The BPS Temperature Equilibration Rate}

For the multi component plasma described in Appendix~\ref{BPSdEdx}, we
assumed that the various components were in thermal equilibrium with
themselves but not with each other. In practice, however, components
exchange energy via Coulomb interactions, and will equilibrate
according to a common temperature according to
\begin{eqnarray}
  \frac{d{\cal E}_{ab}}{dt}
  =   -\, {\cal C}_{ab} \left( T_a - T_b \right) \ .
\label{dedtab}
\end{eqnarray}
As with the stopping power, the rate coefficient can be written as a
sum of three terms,
\begin{eqnarray}
  {\cal C}_{ab}
  =
  \Big({\cal C}^\smC_{ab,\smS} 
  +
  {\cal C}^\smC_{ab,\smR} \Big)
  +
  {\cal C}^\smDelQ_{ab} \ ,
\label{Cab}
\end{eqnarray}
given by (12.31), (12.25), and (12.50) respectively in BPS~\cite{bps}:
\begin{eqnarray}
  {\cal C}^\smC_{ab,\smS} 
  &\!\!=\!& 
  \!-{\kappa_a^2\, \kappa_b^2 } \, 
  \frac{ (\beta_a m_a \beta_b m_b)^{1/2}}{\left( \beta_a m_a + 
  \beta_b m_b \right)^{3/2} } \,\left( \frac{1}{2\pi} \right)^{\!\!3/2}\, 
  \left[\,\ln\!\left\{ \frac{e_a\,e_b}{4 \pi}\,
  \frac{K}{4 \, m_{ab} \, V^2_{ab}}\right\} 
  + 2 \gamma\,  \right] 
\label{classicdone}
\\[10pt]
  {\cal C}^\smC_{ab,\smR} 
  &\!\!=\!&
  \frac{\kappa_a^2\, \kappa_b^2}{2\pi}\!
  \left(\frac{\beta_a m_a}{2\pi} \right)^{\!\! 1/2} \!\!\!
  \left(\frac{\beta_b m_b}{2\pi} \right)^{\!\! 1/2} \!\!\!\!
  \int_{-\infty}^{\infty} \!\! dv \, v^2 
  e^{- \frac{1}{2}( \beta_a m_a + \beta_b m_b) v^2 }  
  \frac{i}{2 \pi} \,\frac{F(v)}{\rho_\text{total}(v)}\, 
  \ln \hskip-0.1cm \left\{ \frac{F(v)}{K^2}\right\} 
\label{nunnn}
\\[10pt]
 {\cal C}_{ab}^\smQ 
  &\!\!=\!&  
  \!-\frac{1}{2} \, \kappa_a^2\, \kappa_b^2 \, 
  \frac{(\beta_a m_a \, \beta_b m_b)^{1/2}}{(\beta_a m_a \!+\! 
  \beta_b m_b)^{3/2}}\! 
  \left(\frac{1}{2\pi}\right)^{\!\!3/2} \!\!\!\!
  \int_0^\infty \!\!\! d \zeta\, e^{-\zeta/2} \!\!
  \left[{\rm Re}\hskip0.02cm 
  \psi\!\left(\!1 + i\frac{\bar\eta_{ab}}{\zeta^{1/2}}\right) \!-\! 
  \ln\!\left\{ \frac{\bar\eta_{ab}}{\zeta^{1/2}}\right\}\!
  \right] .
\label{qqcorr}
\end{eqnarray}
The function $F(v)$ in Eq.~(\ref{nunnn}) is defined by
Eqs.~(\ref{Fdef}) and (\ref{rhototdefA}), and the strength of the
quantum effects associated with the scattering of two plasma species
$a$ and $b$ is characterized by the dimensionless parameter
\begin{eqnarray}
  \bar\eta_{ab} 
  &=& 
  \frac{e_a e_b}{4\pi \hbar V_{ab}}  \ ,
\end{eqnarray}
where the square of the thermal velocity in this expression is defined
by
\begin{eqnarray}
  V_{ab}^2 
  =
  \frac{T_a}{m_a} + \frac{T_b}{m_b} \ .
\end{eqnarray}
In the limit $\beta_e m_e \ll \beta_i m_i$ we have
\begin{eqnarray}
  V_{ei}^2 
  =
  \frac{1}{\beta_e m_e} \ .
\end{eqnarray}
Upon taking the sum over ions, ${\cal C}_{e\smI}={\sum}_i{\cal
C}_{ei}$, and using the inequality $\beta_e m_e \ll \beta_i m_i$,
we can express the rate coefficient as~\cite{bs}
\begin{eqnarray}
  {\cal C}_{e \smI} 
  = 
  \frac{\kappa_e^2}{2\pi}\, 
  \left(\frac{\beta_e m_e}{2\pi}\right)^{1/2} 
  \frac{1}{2}\,{\sum}_i\omega_i^2
  \Bigg[\ln\!\left\{ \frac{8 T_e^2}{\hbar^2 \omega_e^2}\right\} 
  -\gamma -1  - \Delta_i(\bar\eta_{ei})
  \Bigg] \ ,
\label{Ceialleta}
\end{eqnarray}
with
\begin{eqnarray}
  \Delta_i(\bar\eta_{ei})
   = 
  \int_0^\infty \!\! d \zeta\, e^{-\zeta/2} \,
  \Bigg[ {\rm Re}\,\psi\!\left( 1 + i\,
  \frac{\bar\eta_{ei}}{\zeta^{1/2}}\right) + \gamma
  \Bigg] \ .
\label{Deltai}
\end{eqnarray}
In the extreme quantum limit the term $\Delta_i$ in Eq.~(\ref{Deltai})
vanishes, and we obtain Eq.~(\ref{bpsrate}).



\begin{thebibliography}{99}
\bibliographyskip

\bibitem{bps} %
  L.S.~Brown, D.L.~Preston, and R.L.~Singleton~Jr.,
 {\em Charged Particle Motion in a Highly Ionized Plasma}, 
  Phys. Rep. {\bf 410} (2005) 237, 
  arXiv: physics/0501084; For a more detailed pedagogical
  explanation see also Refs.~\cite{lfirst,bs,bpsped}.

\bibitem{lfirst} %
  L.S. Brown, 
  {\em New Use of Dimensional Continuation Illustrated by 
  $dE/dx$ in a Plasma}, 
  Phys. Rev. {\bf D 62} (2000) 045026,
  arXiv: physics/9911056.

\bibitem{bs} L.S. Brown and R.L.~Singleton~Jr, 
  {\em Temperature Equilibration Rate with Fermi-Dirac Statistics}, 
  Phys. Rev. E {\bf 76} (2007) 066404, arXiv: 0707.2370.

\bibitem{bpsped} %
  R.L.~Singleton~Jr., {\em BPS Explained I: Temperature Relaxation in a
  Plasma}, arXiv: 0706.2680; 
  R.L.~Singleton~Jr., 
  {\em BPS Explained II: Calculating the Equilibration Rate
  in the Extreme Quantum Limit}, arXiv: 0712.0639.

\bibitem{fraley74}
  G.S.~Fraley, E.J.~Linnebur, R.J.~Mason, and R.L.~Morse,
  {\em Thermonuclear Burn Characteristics of Compressed
  Deuterium-Tritium Microspheres},
  Phys. Fluids {\bf 17} (1974) 474.

\bibitem{lip}
  C.K. Li and R.D. Petrasso,
  {\em Charged-Particle Stopping Powers in Inertial Confinement 
  Fusion Plasmas},
  Phys. Rev. Lett. {\bf 70} (1993) 3059. 

\bibitem{hoff}
  N.M. Hoffman, private communication (2005).

\bibitem{cor}
  E.G. Corman, W.E. Loewe, G.E. Cooper, and A.M. Winslow,
  {\em Multigroup Diffusion of Energetic Charged Particles},
  Nuc. Fusion {\bf 15} (1975) 377.

\bibitem{leemore}
  Y.T.~Lee and R.M.~More, 
  {\em An Electron Conductivity Model for Dense Plasmas}, 
  Phys. Fluids {\bf 27} (1984) 5. 

\bibitem{hicks00}
  D.G.~Hicks, C.K.~Li, F.H.~S\'{e}guin, A.K.~Ram, J.A.~Frenje, 
  R.D.~Petrasso, 
  J.M.~Soures, V.~Yu.~Glebov, D.D.~Meyerhofer, S.~Roberts, C.~Sorce, 
  C.~St\"{o}ckl, 
  T.C.~Sangster, and T.W.~Phillips, 
  {\em Charged-particle Acceleration and Energy Loss in Laser-Produced
  Plasmas},
  Phys. Plasmas {\bf 7} (2000) 5107; 
  the shot in question is the Omega shot \#16176.

\bibitem{nuc}
  See, for example:
  W.N.~Cottingham and D.A.~Greenwood, {\em An Introduction to 
  Nuclear Physics} (Cambridge University Press, Cambridge, 2001)
  199--206.

\bibitem{gd66}
  H.~Gould and H.~DeWitt, 
  {\em Convergent Kinetic Equation for a Classical Plasma},
  Phys. Rev. {\bf 155} (1967) 68.

\bibitem{by}
  L.~S. Brown and L.~G. Yaffe, 
  {\em Effective Field Theory for Highly Ionized Plasmas}, 
  Phys. Rep. {\bf 340} (2001) 1-164, 
  arXiv: physics/9911055. 

%
%
%
%




\end{thebibliography}
\end{document}